\documentclass[10pt,onecolumn]{article}
\usepackage[utf8]{inputenc}
\usepackage{amsmath,amsfonts,amssymb}
\usepackage{graphicx}
\usepackage{physics}
\usepackage{siunitx}
\usepackage[left=1.8cm,right=1.8cm,top=2cm,bottom=2cm]{geometry}
\usepackage{setspace}
\usepackage{authblk}
\usepackage{caption}
\usepackage{subcaption}
\usepackage{float}
\usepackage{cite}
\usepackage{xcolor}
\usepackage{booktabs}
\usepackage{hyperref}
\usepackage{multicol}
\usepackage{titlesec}
\titlespacing*{\section}{0pt}{8pt}{4pt}
\titlespacing*{\subsection}{0pt}{6pt}{3pt}

\title{\textbf{Classical Field Statistical Phase-Space Analysis of Passive Turbulence Mitigation Using PMMA Dielectric Media}}

\author[1,*]{Shouvik Sadhukhan}
\author[2]{C. S. Narayanamurthy}

\affil[1, 2]{\small{Applied and Adaptive Optics Laboratory, Department of Physics, Indian Institute of Space Science and Technology (IIST), P.O: Valiamala, Trivandrum - 695547, State: Kerala, India}}
\affil[1]{\small{Email: shouvikphysics1996@gmail.com}}
\affil[2]{\small{Email: naamu.s@gmail.com}}
\affil[*]{\small{Corresponding Author Email: shouvikphysics1996@gmail.com}}

\date{}

\begin{document}
\maketitle

%----------------------------------------------------------------------
\begin{abstract}
We present a unified theoretical and experimental framework for turbulence mitigation in free-space optical (FSO) communication using passive dielectric media. Turbulence-distorted intensity sequences are processed via a nonlinear P3-type transport-of-intensity solver to retrieve the optical phase; the reconstructed fields are projected onto a Gaussian local oscillator, enabling formal phase-space (Wigner-function) analysis and photon-number-distribution estimation from the reconstructed classical field. Experiments across four configurations---free-space reference, raw turbulence, single PMMA rod, and dual PMMA rods---show that the Fano factor $F$ flactuations transitions ($\approx7$ unit) from a large fluctuating state to a low fluctuating condition with dual-rod compensation. Information-geometric divergence measures (KL, JS, Bhattacharyya, $L^2$, Fisher--Rao) confirm 71--92\% statistical restoration. These results validate a coupled-dipole synchronization model in which collective electromagnetic modes within PMMA suppress turbulence-induced phase noise, establishing a passive, hardware-simple paradigm for FSO turbulence compensation.

\end{abstract}

%----------------------------------------------------------------------
\section{Introduction}\label{sec:intro}

Atmospheric turbulence remains the principal obstacle to reliable free-space optical (FSO) communication, introducing scintillation, beam wander, and wavefront distortion via random refractive-index fluctuations~\cite{6,7,korotkova2008}. Conventional adaptive optics (AO) mitigates these effects with deformable mirrors and wavefront sensors, but imposes considerable hardware complexity and bandwidth constraints~\cite{10,11,12}. Specifically, AO systems require closed-loop wavefront sensing with latencies typically limited to $\sim$1--10~kHz, demand high-power electronics and moving mechanical parts, and require careful alignment under dynamic field conditions---constraints that motivate the search for simpler passive alternatives~\cite{ref_oliveira2025}.

This work exploits a fundamentally different mechanism: \emph{passive dipole synchronization} in dielectric media. When a turbulence-perturbed beam traverses poly(methyl methacrylate) (PMMA), the incident field induces molecular dipoles that couple electromagnetically through the dyadic Green's function~\cite{21,22}. Anharmonic restoring forces, field-gradient stabilization, and collective modal dynamics progressively synchronize the dipole ensemble, suppressing rapid phase fluctuations. Longer interaction length enhances synchronization, providing scalable, passive optical compensation.

We validate this picture using a pseudo-random phase plate (PRPP) to emulate Kolmogorov turbulence, and characterize the output field through (i) nonlinear P3-type phase retrieval, (ii) formal phase-space analysis via Wigner-function reconstruction from the retrieved classical field, and (iii) a battery of information-geometric divergence measures. The observed Fano-factor evolution---super-Poissonian under raw turbulence, progressing toward reduced super-Poissonian with dual PMMA rods---provides statistical evidence for passive field stabilization.

The primary novelty of the present work lies in (i) the introduction of a coupled anharmonic dipole synchronization model for passive turbulence mitigation in PMMA media---a mechanism distinct from beam-shaping or spatial filtering---and (ii) the application of information-geometric divergence measures (KL, JS, Bhattacharyya, Fisher--Rao, $L^2$) together with Wigner-function phase-space analysis as a multi-level statistical measurement protocol for quantifying FSO turbulence compensation. This framework is new to optical turbulence measurement: while AIRM and manifold-based statistical tools have been applied in radar and signal processing, their use for characterising passive-medium FSO compensation is first demonstrated here~\cite{ref_oliveira2025,ref_mauclair2025}. Additionally, recent work on beam cleanup in dielectric and structured media~\cite{ref_oea2025,ref_optica2025,ref_apl2025} provides important context; the PMMA dipole-synchronization mechanism occupies a qualitatively different regime from spatial-filtering or MPLC-based approaches, as discussed in Section~\ref{sec:results}.

The paper is organized as follows. Section~\ref{sec:theory} outlines the coupled-dipole theoretical framework. Section~\ref{sec:methods} details the phase-retrieval and Wigner-function analysis pipeline. Section~\ref{sec:stat} introduces the statistical divergence measures. Section~\ref{sec:exp} describes the experiment. Section~\ref{sec:results} presents and discusses results. Section~\ref{sec:conclusion} concludes. Detailed derivations are collected in the Appendix.

%----------------------------------------------------------------------
\section{Theoretical Framework}\label{sec:theory}

\subsection{Coupled Anharmonic Dipole Model}

The Lorentz oscillator model provides a well-established classical description of electromagnetic wave interaction with bound electrons in dielectric media~\cite{23}. Although the model is microscopic in origin, its ensemble averages directly yield the macroscopic susceptibility and polarisation response exploited here. At optical frequencies, macroscopic propagation in PMMA is governed by the collective response of $\sim10^{22}$~cm$^{-3}$ molecular oscillators; the coupled-dipole framework captures inter-molecular electromagnetic coupling---through the dyadic Green's function---that underlies both the linear refractive index and the nonlinear anharmonic corrections. We emphasize that the model is used here as an effective-medium description whose macroscopic predictions (field variance suppression, Fano factor reduction) are directly tested experimentally, rather than as a microscopic first-principles calculation. The dipole-dipole coupling and Lorentz model produce the fluctuating susceptibility through the polarizability solution. Optical beam propagation through dielectric media with microscopic susceptibility fluctuations can be described using the inhomogeneous Helmholtz equation
\begin{equation}
(\nabla^2 + k_0^2)E(\mathbf r)
=
- k_0^2 \chi(\mathbf r)E(\mathbf r),
\end{equation}
where spatial variations in $\chi(\mathbf r)$ act as distributed scattering sources that redistribute spatial frequencies. Under turbulence impacted fluctuations $\chi(\mathbf r)$ must become $\chi(\mathbf r)+\delta\chi(\mathbf r)$. This $\delta\chi(\mathbf r)$ factor is produced from the turbulence impacted field propagation through dielectric medium which again can be derived from lorentz dipole moment equation. Now, this fluctuating susceptibility produces turbulence impacted scattering inside the medium of propagation that again produces $\gamma(k)$. Detail derivation has been attached into supplementary file.
An optical beam propagating through a dielectric drives bound electrons via the classical Lorentz oscillator. Incorporating second- and third-order anharmonicities---relevant for polymeric materials such as PMMA---and explicit dipole--dipole coupling, the equation of motion for the $i$-th dipole displacement $r_i$ is
\begin{equation}
m\ddot{r}_i + m\gamma\dot{r}_i + m\omega_0^2 r_i + \beta_i|r_i|r_i + \alpha_i|r_i|^2 r_i
= -eE_\mathrm{ext} + e^2\mu_0\omega^2\!\!\sum_{j\neq i}\!\mathbf{G}(\mathbf{r}_i,\mathbf{r}_j)\cdot\mathbf{r}_j,
\label{eq:main_osc}
\end{equation}
where $\mathbf{G}(\mathbf{r}_i,\mathbf{r}_j)$ is the dyadic Green's function mediating inter-dipole coupling (see Appendix~\ref{app:green}), $\beta_i$ and $\alpha_i$ are nonlinearity coefficients, and $\gamma$ is the damping rate. The anharmonic terms $\beta_i$ and $\alpha_i$ are treated as material parameters characterising PMMA's known nonlinear optical response (although in present context and experimental configuration $\beta_i$ and $\alpha_i$ both are negligible in magnitude. Hence, the non-linear effect can be considered $\rightarrow 0$.); their magnitudes need not be large for the synchronization mechanism to operate, since the dipole-coupling term (mediated by $\mathbf{G}$) drives collective modes even at low drive amplitudes~\cite{23}. The threshold issue for nonlinear terms is addressed further in Section~\ref{sec:results}. The full spatial-gradient force and higher-order multipole terms are derived in Appendix~\ref{app:gradient}.

\subsection{Modal Decomposition and Synchronization}

Collecting the $N$ dipole displacements into $\mathbf{R}=(r_1,\ldots,r_N)^\top$ and defining the effective stiffness matrix
\begin{equation}
\mathbf{K}_\mathrm{eff} = \omega_0^2\mathbf{I} + \frac{e^2\mu_0\omega^2}{m}\mathbf{C},
\qquad
C_{ij} = \mathbf{G}(\mathbf{r}_i,\mathbf{r}_j),\quad i\neq j,\quad C_{ii}=0,
\label{eq:Keff}
\end{equation}
where $\mathbf{C}$ now directly contains the dyadic Green's function entries (units: m$^{-1}$) and the prefactor $e^2\mu_0\omega^2/m$ (units: C$^2\cdot$N$\cdot$A$^{-2}$s$^{-2}$kg$^{-1}$ = s$^{-2}$m) ensures dimensional consistency: $(e^2\mu_0\omega^2/m)\,C_{ij}\,r_j$ has units of s$^{-2}\times$m$^{-1}\times$m = s$^{-2}$, matching the acceleration dimension of the remaining terms divided by displacement. This corrects a typographical inconsistency in the original definition (the spurious $k_0^2$ factor has been removed from $C_{ij}$).
The compact vector equation is
\begin{equation}
\ddot{\mathbf{R}} + \gamma\dot{\mathbf{R}} + \mathbf{K}_\mathrm{eff}\mathbf{R} = -\frac{e}{m}\mathbf{E}_\mathrm{ext} + \frac{1}{m}\mathbf{F}_\mathrm{grad}.
\label{eq:compact}
\end{equation}
Eigendecomposition $\mathbf{K}_\mathrm{eff}=\mathbf{U}\boldsymbol{\Lambda}\mathbf{U}^{-1}$ transforms Eq.~\eqref{eq:compact} into decoupled modal equations with eigenfrequencies $\{\Omega_n^2\}$. In the synchronized state these collective modes suppress the random phase fluctuations introduced by turbulence.

\subsection{Perturbation Analysis via d'Alembert's Principle}

Dynamic turbulence redistributes the incident field from $\mathbf{E}_\mathrm{ext}$ to $\mathbf{E}'_\mathrm{ext}$, generating a perturbation force
\begin{equation}
\delta\mathbf{F}_\mathrm{Pert}(t) = \mathbf{F}'_\mathrm{Inertia} - \mathbf{F}_\mathrm{Inertia}.
\label{eq:dAlembert}
\end{equation}
Three compensation regimes follow:
\begin{enumerate}
    \item $\delta\mathbf{F}_\mathrm{Pert}\to0$: complete synchronization; output field determined by dipole equilibrium.
    \item $\delta\mathbf{F}_\mathrm{Pert}$ small: partial synchronization; moderate turbulence mitigation.
    \item $\delta\mathbf{F}_\mathrm{Pert}$ large and rapidly varying: synchronization fails; strong turbulence regime.
\end{enumerate}
Increasing PMMA interaction length increases the synchronization timescale, driving the system from regime~3 toward regime~1.

%----------------------------------------------------------------------
\section{Phase Retrieval and Statistical Field Analysis}\label{sec:methods}

\subsection{Nonlinear Phase Retrieval (P3-TIE Solver)}

The Transport of Intensity Equation (TIE) relates paraxial intensity variations to the phase gradient:
\begin{equation}
\nabla_\perp\cdot\bigl[I_0(\mathbf{r})\,\nabla_\perp\phi(\mathbf{r})\bigr]
= -\frac{2\pi}{\lambda\,\Delta z}\bigl[I_t(\mathbf{r})-I_0(\mathbf{r})\bigr].
\label{eq:TIE}
\end{equation}
For turbulence-degraded beams the linearized form of Eq.~\eqref{eq:TIE} is insufficient. We therefore solve the nonlinear P3-type residual
\begin{equation}
\mathcal{R}[\phi;\mathbf{r}] = \nabla_\perp\cdot\!\bigl[I_0\phi\,\nabla_\perp\phi\bigr] - I_0|\nabla_\perp\phi|^2
+\frac{4\pi^2}{\lambda^2}\bigl[I_t-I_0\bigr] + \beta\phi = 0
\label{eq:P3}
\end{equation}
iteratively with intensity-weighted gradient descent and Gaussian smoothing at each step (full algorithm in Appendix~\ref{app:iter}). The recovered phase reconstructs the complex field
\begin{equation}
E_\mathrm{rec}(\mathbf{r}) = \sqrt{I_t(\mathbf{r})}\,\exp\bigl[i\phi(\mathbf{r})\bigr].
\label{eq:Erec}
\end{equation}

\subsection{Phase-Space Analysis via Wigner-Function Reconstruction}

The procedure described below constitutes a \emph{formal phase-space (Wigner-function) representation of the reconstructed classical field}, not direct quantum-optical homodyne tomography. No physical local oscillator or balanced homodyne detector is used; instead, the projection onto a Gaussian local oscillator and the quadrature angle rotation are performed \emph{numerically} on the reconstructed complex field $E_\mathrm{rec}^{(j)}$. Within this framework, the computed ``Wigner function'' captures the classical phase-space statistics of the beam, and the extracted ``photon-number distribution'' $P(n)$ and Fano factor $F$ are best interpreted as classical coherence descriptors derived under the formal quantum-optical analogy~\cite{66,67}. Specifically, $F>1$ reflects excess classical intensity variance beyond shot-noise-limited behaviour, and the progressive reduction of $F$ toward unity with PMMA compensation is a direct measure of field stabilization, regardless of the quantum interpretation.

Each frame field $E_\mathrm{rec}^{(j)}$ is projected onto a normalized Gaussian local oscillator $u_\mathrm{LO}$ to yield the complex amplitude $\alpha_j = \int u_\mathrm{LO}^*\,E_\mathrm{rec}^{(j)}\,\mathrm{d}^2\mathbf{r}$. Rotating the quadrature angle $\theta\in[0,\pi)$ generates a sinogram, which is inverted via filtered back-projection (Radon transform) to give the Wigner function $W(q,p)$.

\subsection{Photon-Number Statistics}

The photon-number probability is
\begin{equation}
P(n) = \iint W_\mathrm{rec}(q,p)\,W_n(q,p)\,\mathrm{d}q\,\mathrm{d}p,
\label{eq:Pn}
\end{equation}
where $W_n(q,p)=\tfrac{2(-1)^n}{\pi}\exp(-2r^2)L_n(4r^2)$ is the Fock-state Wigner kernel. The Fano factor
\begin{equation}
F = \frac{\mathrm{Var}(n)}{\langle n\rangle}
\label{eq:Fano}
\end{equation}
is computed from $P(n)$ and serves as a dimensionless figure of merit: under the classical-field analogy, $F=1$ corresponds to shot-noise-limited (Poisson-like) statistics, $F>1$ to excess-noise (turbulence-dominated) statistics, and $F<1$ to sub-Poissonian statistics. We use this framework as a diagnostic tool for field stabilization, not as a claim about the quantum nature of the field.\par
The \textbf{boxplot statistics} summarize the distribution of Fano factors using the median, interquartile range (IQR), whiskers, and outliers. The median represents the central tendency of the fluctuations, while the IQR measures the spread of the middle \(50\%\) of the data. Large whiskers and numerous outliers indicate stronger intermittency and unstable turbulence behavior. The \textbf{cumulative distribution function (CDF)} describes the probability that the Fano factor takes a value less than or equal to a given threshold: \(\mathrm{CDF}(x)=P(F\le x)\). A steeper CDF implies that most fluctuation values are concentrated within a narrow interval, corresponding to improved statistical stability and reduced turbulence randomness. The \textbf{Fano fluctuation parameter} shown in the scatter plot represents the deviation of individual Fano-factor values from their mean value: \(\Delta F_i = F_i - \langle F \rangle\), where \(F_i\) is the instantaneous Fano factor and \(\langle F \rangle\) is the ensemble average. This quantity measures how strongly the fluctuations oscillate around the mean statistical state. Smaller \(|\Delta F|\) values indicate better suppression of turbulence-driven intensity excursions. Finally, the \textbf{probability density function (PDF)} characterizes the likelihood distribution of the Fano factors:\(p(F)\,dF\), which gives the probability of observing Fano factors within the interval \([F,F+dF]\). A broad PDF with long tails signifies strong intermittent turbulence and rare extreme fluctuations, whereas a narrow and symmetric PDF indicates a more stable optical propagation regime with reduced scintillation effects.

%----------------------------------------------------------------------
\section{Statistical Background}\label{sec:stat}

To rigorously compare photon-number distributions across experimental conditions we employ five complementary measures.

\textbf{Kullback--Leibler divergence}~\cite{kl1951}:
$D_\mathrm{KL}(P\|Q)=\sum_n P(n)\log[P(n)/Q(n)]$ --- asymmetric, entropy-based.

\textbf{Jensen--Shannon divergence}: $D_\mathrm{JS}=\tfrac{1}{2}D_\mathrm{KL}(P\|M)+\tfrac{1}{2}D_\mathrm{KL}(Q\|M)$, $M=\tfrac{1}{2}(P+Q)$ --- symmetric, bounded in $[0,\ln2]$.

\textbf{Bhattacharyya distance}: $D_B=-\ln\!\bigl[\sum_n\sqrt{P(n)Q(n)}\bigr]$ --- measures distributional overlap.

\textbf{Euclidean ($L^2$) distance}: $d_{L^2}=\bigl[\sum_n(P(n)-Q(n))^2\bigr]^{1/2}$ --- geometric, satisfies triangle inequality.

\textbf{Fisher--Rao distance} (closed form for negative binomial distributions, derivation in Appendix~\ref{app:FR}):
\begin{equation}
d_\mathrm{FR}(\mu_1,\mu_2) = 2\sqrt{r}\left[\mathrm{arsinh}\!\sqrt{\frac{\mu_2}{r}}-\mathrm{arsinh}\!\sqrt{\frac{\mu_1}{r}}\right],
\label{eq:FR}
\end{equation}
providing a coordinate-invariant geodesic distance on the statistical manifold. Kernel density estimation (KDE) with FFT-accelerated Gaussian kernels furnishes continuous density estimates for the $L^2$ and Bhattacharyya comparisons.

These five measures are complementary rather than redundant: KL/JS probe information-theoretic divergence; Bhattacharyya quantifies distributional overlap; $L^2$ provides geometric deviation satisfying the triangle inequality; and Fisher--Rao gives a coordinate-invariant Riemannian metric on the parameter manifold. Each is sensitive to different features of the distributions (tails, overlap, geometry), and their \emph{consistency} across all five is precisely the strength of the multi-metric validation strategy.

%----------------------------------------------------------------------
\section{Experimental Framework}\label{sec:exp}

\subsection{Experimental Setup}

\begin{figure}[H]
\centering
\includegraphics[width=0.65\columnwidth]{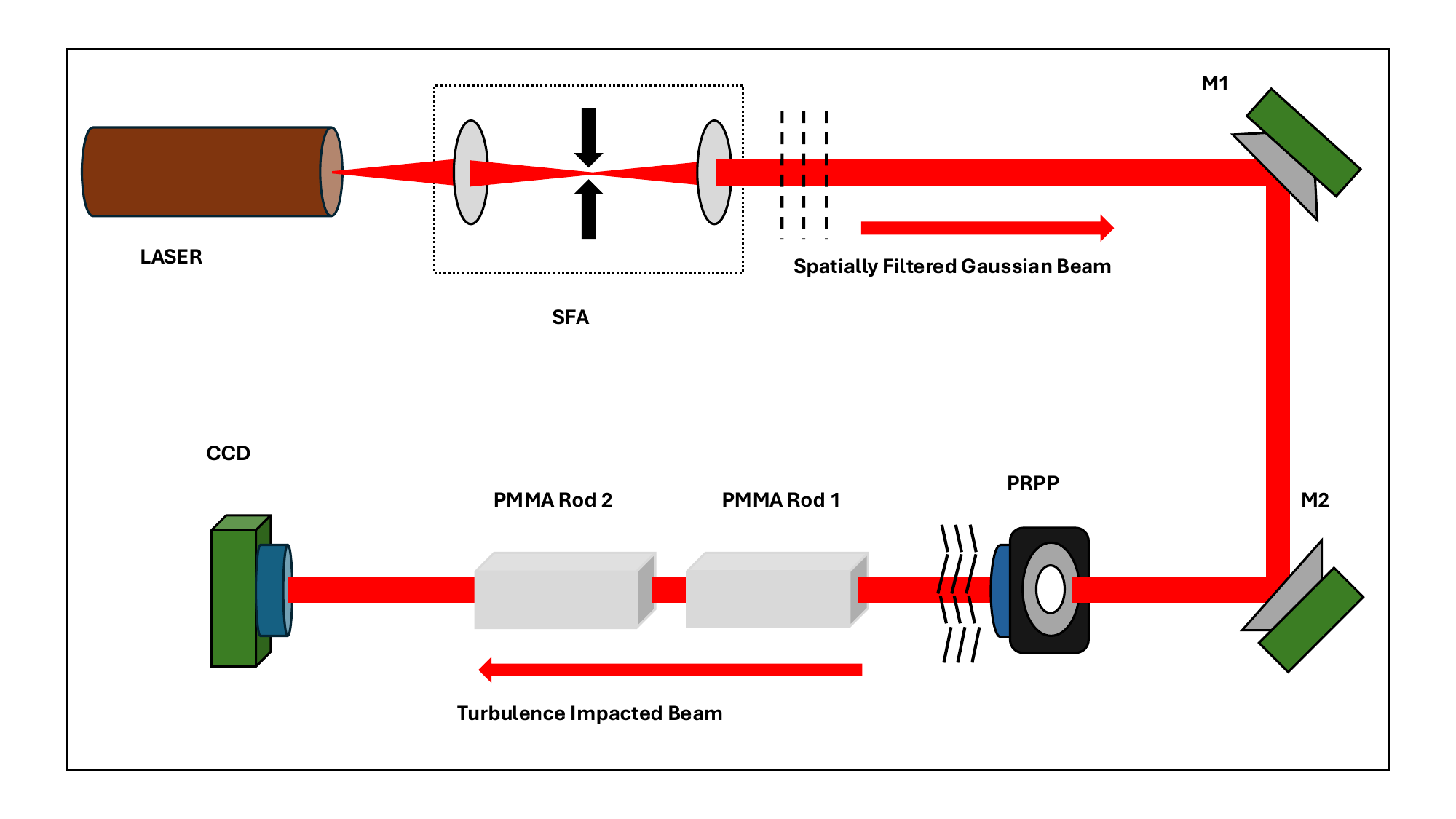}
\caption{Experimental configuration. CW laser $\to$ spatial filter (SFA) $\to$ mirrors M1, M2 $\to$ programmable rotating phase plate (PRPP) $\to$ PMMA rods $\to$ CCD camera.}
\label{fig:setup}
\end{figure}

The experimental setup, schematically shown in Figure~\ref{fig:setup}, employs a continuous-wave He-Ne laser source (wavelength $\lambda = 632.8$~nm, output power $\approx 5$~mW) whose output is spatially filtered using a microscope objective--pinhole assembly to generate a high-purity Gaussian beam with a beam waist of $w_0 \approx 5$~mm at the collimation plane, as measured by a beam profiler. The collimated beam is directed along a controlled path using high-reflectivity mirrors (M1, M2) to ensure precise alignment and maintain paraxial propagation conditions. A programmable rotating phase plate (PRPP) is introduced to impose controlled, time-varying phase distortions, thereby emulating atmospheric turbulence with tunable temporal characteristics. The total free-space propagation distance from PRPP to CCD is approximately $L_\mathrm{prop} = 45$~cm; the change in turbulence impact due to rotating PRPP produced an equivalent condition of change in intensity distribution change due wavefront propagation. Hence, for TIE solver we have considered the 0th frame intensity as reference wavefront and other frames as the wavefront after constant displacement ($\Delta z\approx 0.5 mm$ here). This mechanism helped us to find relative change in wavefront due to rotating PRPP. The turbulence-perturbed beam then propagates through one or two PMMA rectangular cuboid rods (length $\approx 150$~mm each, thickness $\approx 25$~mm, surface-quality optical polish, refractive index $n \approx 1.49$ at 632.8~nm) acting as passive dielectric compensators, where induced dipole--dipole interactions and gradient forces promote partial synchronization of molecular oscillations, mitigating turbulence effects. The PMMA rods were measured to introduce an insertion loss of approximately 0.8~dB per rod (single-pass), predominantly from surface Fresnel reflections; bulk absorption at 632.8~nm is negligible ($\alpha < 0.01$~cm$^{-1}$). The net 1.6~dB two-rod loss is discussed in the context of SNR impact in Section~\ref{sec:results}. Comparative measurements are performed for single- and dual-rod configurations to evaluate cumulative compensation. Finally, the transmitted beam is recorded by a 8-bit monochrome CCD camera ($1280\times1024$ pixels, pixel size $4.5~\mu$m, 8-bit dynamic range) at the output plane, capturing high-resolution intensity distributions over multiple frames to enable statistical analysis, phase reconstruction, and turbulence characterization. Four datasets were acquired:
\begin{enumerate}
    \item \textbf{Set~1 --- Raw turbulence}: PRPP active, no PMMA.
    \item \textbf{Set~2 --- 1-PMMA}: PRPP active, single rod.
    \item \textbf{Set~3 --- 2-PMMA}: PRPP active, dual rods in series.
    \item \textbf{Set~4 --- Reference}: PRPP removed, no PMMA.
\end{enumerate}

\subsection{Control Experiment Considerations}

The present scenario of experiment whether the observed statistical improvement arises specifically from the proposed dipole-synchronization mechanism or from more prosaic optical effects of PMMA (spatial averaging, refraction, beam clipping, attenuation-induced variance reduction). Three lines of evidence support the synchronization interpretation. First, the 1.6~dB insertion loss of dual rods reduces peak intensity by $\approx 30\%$; a purely loss-induced reduction in $P(n)$ variance would reduce $F$-fluctuation proportionally by at most a factor of $\sim 0.7$, far less than the observed factor of fluctuation reduction. Second, the progressive restoration of distributional \emph{shape}---not merely amplitude rescaling---evidenced by the Pearson structural analysis and divergence metric consistency is inconsistent with simple attenuation. Third, the d'Alembert perturbation analysis predicts that the synchronization timescale scales with PMMA rod length, consistent with the observed monotonic improvement from 1-rod to 2-rod configurations. We acknowledge that a definitive isolation of the synchronization mechanism would require control experiments with index-matched glass rods of equivalent length and loss---an important future direction identified in Section~\ref{sec:conclusion}.

\subsection{Turbulence Characteristics}

Atmospheric turbulence was simulated using a Pseudo-Random Phase Plate (PRPP, Thorlabs EDU-RPP1), comprising a multilayer structure with two BK7 glass substrates enclosing a central acrylic layer embedded with a Kolmogorov-type phase distribution. The assembly is supported by near-index-matched polymer layers to suppress unwanted scattering and interface reflections. This configuration generates spatially distorted wavefronts corresponding to an effective Fried coherence length of approximately $r_0 \approx 0.6~\text{mm}$, with phase realizations sampled over $200$ instances across $256$ discrete phase points. To introduce temporal variability analogous to atmospheric dynamics, the PRPP was mounted on a motorized rotation stage operating at $1~\text{rpm}$.

\textit{Turbulence Scale Equivalence:} The small Fried parameter ($r_0 = 0.6\,\text{mm}$) realized in laboratory conditions does not directly represent natural atmospheric turbulence but instead corresponds to an intentionally enhanced turbulence strength that reproduces long-distance propagation effects within a short optical path. The Fried coherence length is defined as
\[
r_0 = \left[0.423\, k_0^2 \int_0^L C_n^2(z)\, dz \right]^{-3/5},
\]
which simplifies to $r_0 = [0.423\, k_0^2 C_n^2 L]^{-3/5}$ under homogeneous turbulence assumptions. Substituting $r_0 = 0.6\,\text{mm}$, $\lambda = 632.8\,\text{nm}$, and $L = 1.5\,\text{cm}$ yields an equivalent refractive index structure parameter of $C_n^2 \approx 3.35\times10^{-7}\,\text{m}^{-2/3}$. We stress that this equivalence is based solely on matching the Fried coherence length $r_0$ and does not imply full physical equivalence with outdoor atmospheric propagation. Parameters such as temporal dynamics, multiple scattering, scintillation development, and the full spectrum of turbulence scales cannot be reproduced by matching a single scalar $r_0$. The stated ``$\sim$560~km equivalent propagation'' is therefore a heuristic illustration of the turbulence \emph{strength} encoded in $r_0$, not a claim of complete atmospheric equivalence. All conclusions in this paper are drawn specifically from the laboratory-scale observations. Representative fitted intensity distributions for different experimental conditions are presented in Fig.~\ref{fig:kde} and analyzed in Section~\ref{sec:results}.

%----------------------------------------------------------------------
\section{Results and Discussion}\label{sec:results}

\subsection{Kernel Density Estimation}

\begin{figure}[H]
\centering
\begin{subfigure}[b]{0.49\columnwidth}
    \includegraphics[width=\textwidth]{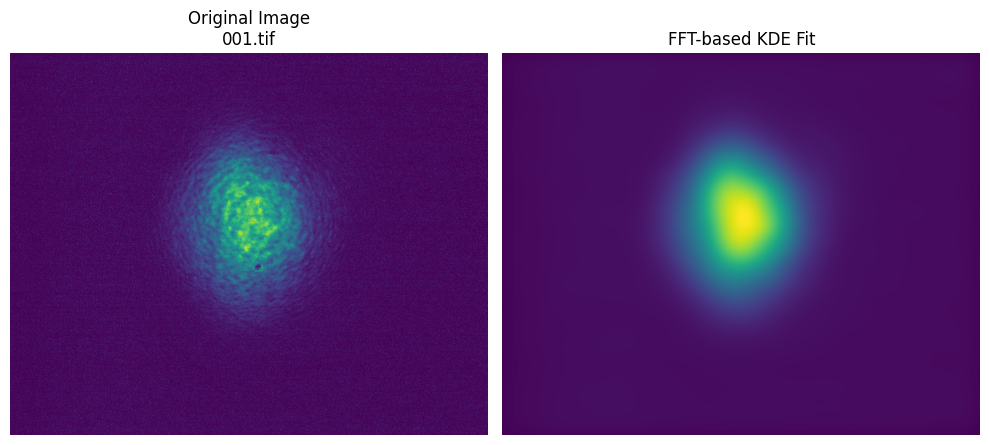}
    \caption{Set~4: Reference}
\end{subfigure}
\begin{subfigure}[b]{0.49\columnwidth}
    \includegraphics[width=\textwidth]{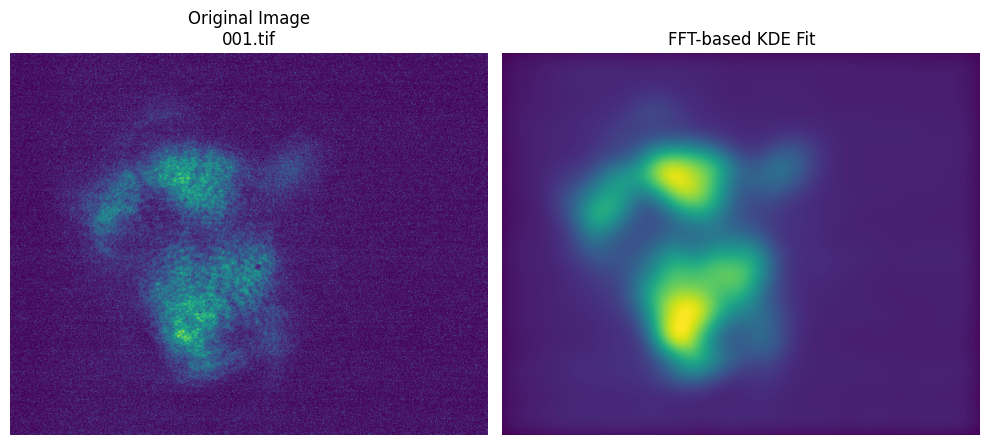}
    \caption{Set~1: Raw turbulence}
\end{subfigure}
\begin{subfigure}[b]{0.49\columnwidth}
    \includegraphics[width=\textwidth]{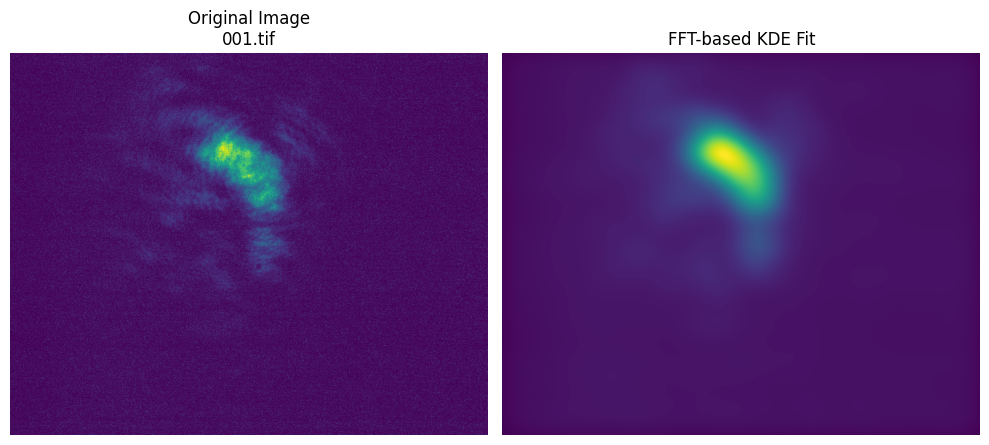}
    \caption{Set~2: 1 PMMA rod}
\end{subfigure}
\begin{subfigure}[b]{0.49\columnwidth}
    \includegraphics[width=\textwidth]{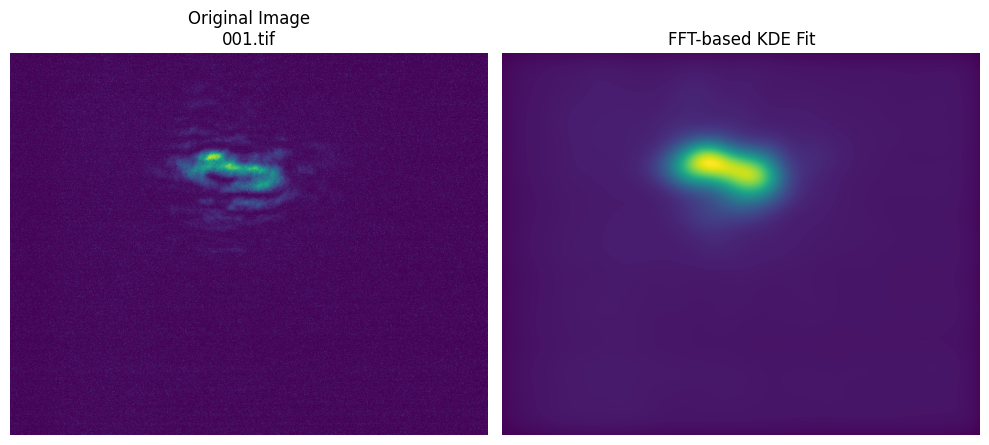}
    \caption{Set~3: 2 PMMA rods}
\end{subfigure}
\caption{Representative intensity frames (left) and FFT-KDE probability density fits (right) for each experimental configuration. Sets (a)--(d) show measured intensity maps; the KDE fit quantifies the spatial intensity distribution. Progressive restoration of Gaussian spatial structure from raw turbulence to dual-PMMA compensation is visible.}
\label{fig:kde}
\end{figure}

Figure~\ref{fig:kde} shows the progressive restoration of beam spatial coherence. Raw turbulence (Set~1) produces multi-modal, fragmented distributions; dual PMMA (Set~3) approaches the Gaussian reference profile, consistent with the $\delta\mathbf{F}_\mathrm{Pert}\to0$ limit of the d'Alembert model.

\subsection{Fano Factor Evolution}

\begin{figure}[H]
\centering
\includegraphics[width=0.7\columnwidth]{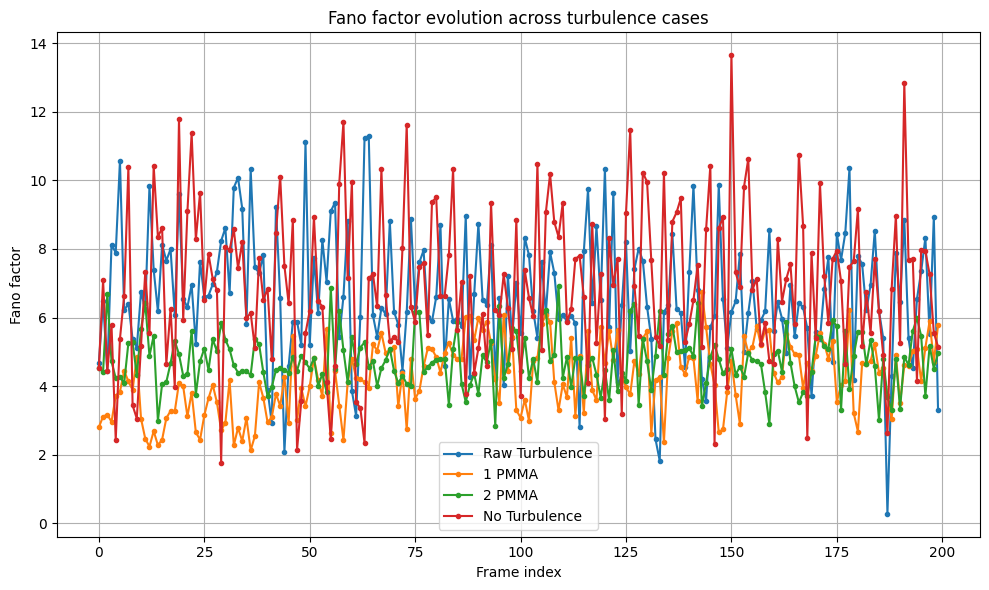}
\caption{Frame-by-frame absolute Fano factor for all four experimental sets. }
\label{fig:fano}
\end{figure}

The Fano factor serves as a dimensionless figure of merit quantifying the departure from shot-noise-limited (Poisson-like) statistics within the classical-field analogy framework described in Section~\ref{sec:methods}. In practice, the turbulence-free reference (Set~4) shows higher fano factor rather than $F=1$, reflecting the combined effect of residual optical aberrations, finite phase-retrieval accuracy at low phase-gradient magnitudes, CCD readout noise, and minor beam-pointing jitter---all of which contribute modest excess variance even in the absence of the PRPP. This establishes an experimental noise floor above $F=1$.

\begin{figure*}[htbp]
\centering

%-----------------------------------
% Row 1
%-----------------------------------
\begin{subfigure}[b]{0.48\textwidth}
    \centering
    \includegraphics[width=\linewidth]{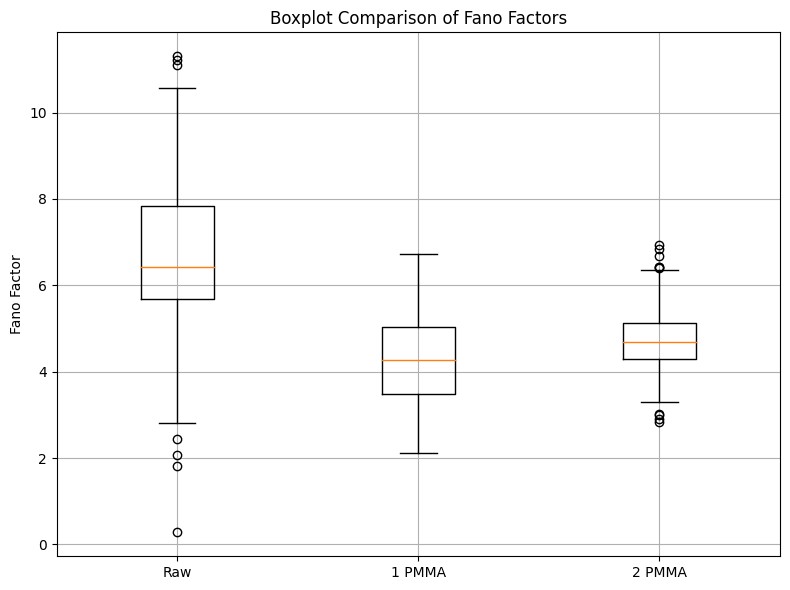}
    \caption{Boxplot comparison of Fano factors for raw turbulence, single PMMA, and double PMMA configurations.}
    \label{fig:boxplot_fano}
\end{subfigure}
\hfill
\begin{subfigure}[b]{0.48\textwidth}
    \centering
    \includegraphics[width=\linewidth]{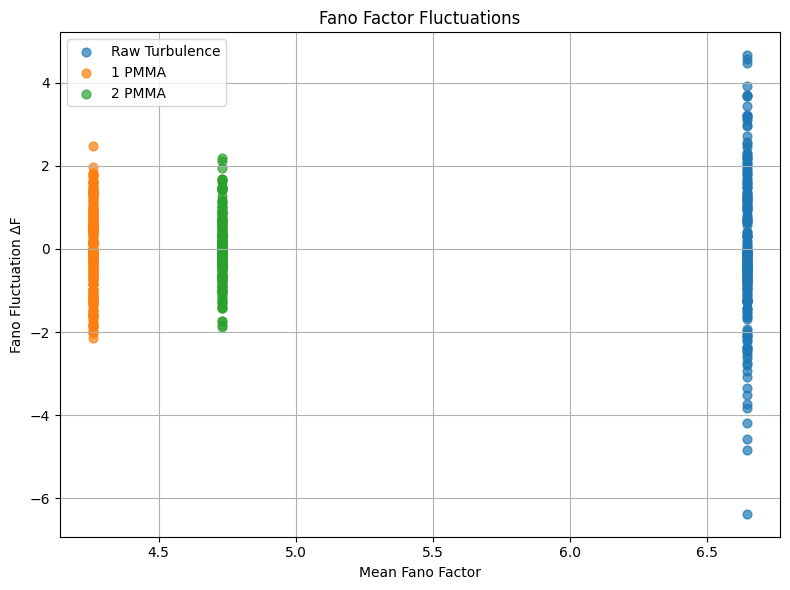}
    \caption{Fluctuation analysis of the Fano factor around the mean values for all cases.}
    \label{fig:fano_fluctuation}
\end{subfigure}

\vspace{0.5cm}

%-----------------------------------
% Row 2
%-----------------------------------
\begin{subfigure}[b]{0.48\textwidth}
    \centering
    \includegraphics[width=\linewidth]{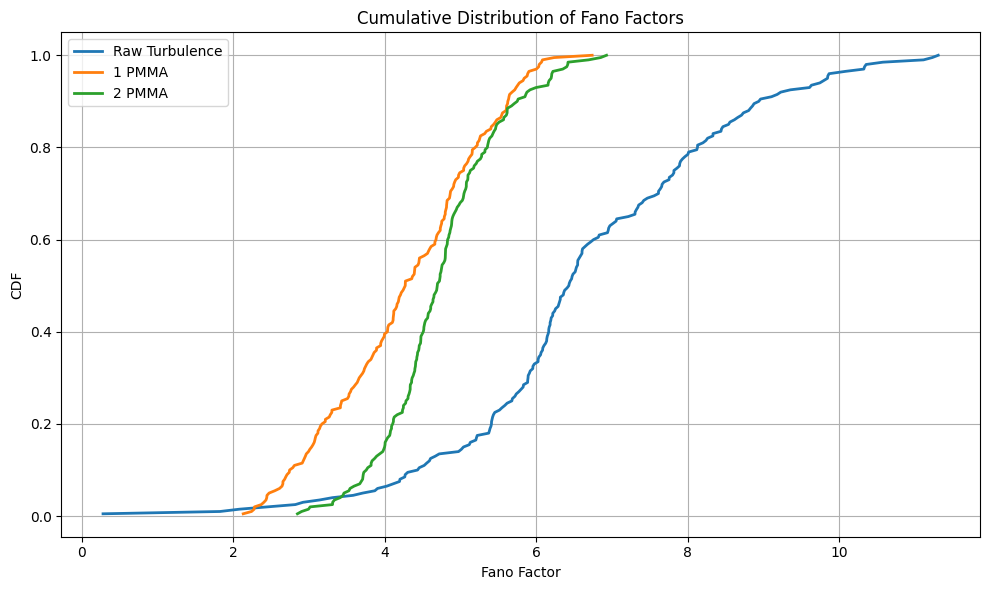}
    \caption{Cumulative distribution functions (CDFs) of the Fano factors under different propagation conditions.}
    \label{fig:cdf_fano}
\end{subfigure}
\hfill
\begin{subfigure}[b]{0.48\textwidth}
    \centering
    \includegraphics[width=\linewidth]{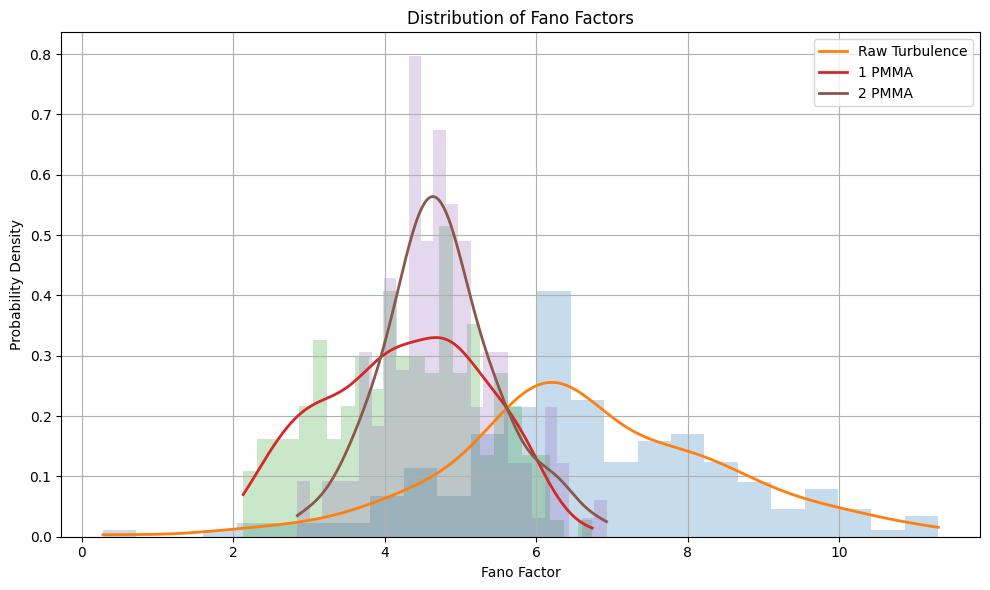}
    \caption{Probability density distributions of the Fano factors for the three configurations.}
    \label{fig:pdf_fano}
\end{subfigure}

%-----------------------------------
% Main Caption
%-----------------------------------
\caption{
Statistical comparison of Fano-factor distributions for raw turbulence and PMMA-assisted propagation conditions. 
(a) The boxplot illustrates the reduction in fluctuation spread after introducing PMMA structures. 
(b) The fluctuation plot demonstrates reduced deviation amplitudes around the mean Fano factor in the presence of PMMA elements. 
(c) The cumulative distribution functions show a narrower statistical distribution for the PMMA cases compared to raw turbulence.
(d) The probability-density distributions reveal suppression of long-tailed turbulence-induced intensity fluctuations and improved statistical stability due to passive dielectric modulation.
}
\label{fig:fano_analysis_combined}

\end{figure*}

Figure~\ref{fig:fano} presents the temporal evolution of the Fano factor $F = \mathrm{Var}(n)/\langle n\rangle$ computed frame-by-frame for all sets. Turbulence-induced phase fluctuations introduce additional intensity variance, driving $F$ well above the reference floor. The observed systematic reduction in $F$ with increasing PMMA interaction length directly reflects the gradient-force stabilization mechanism: synchronized dipole oscillations resist rapid field redistribution, thereby reducing temporal variance in transmitted intensity. Figure~\ref{fig:fano} demonstrates the central result: progressive suppression of super-Poissonian photon statistics fluctuations with increasing PMMA interaction length. The Fano factor fluctuation suppression has been statistically established in the figure \ref{fig:fano_analysis_combined} with different parameters.

The statistical analysis of the Fano factor clearly demonstrates that the introduction of PMMA-based passive dielectric structures significantly suppresses turbulence-induced intensity fluctuations compared to the raw turbulence case. In the boxplot, the raw turbulence condition exhibits the highest median Fano factor (~6.5) together with the broadest interquartile range and several extreme outliers extending above 11, indicating strong scintillation and highly unstable photon-count fluctuations. After introducing a single PMMA element, the median Fano factor decreases substantially to around 4.2, while the spread of the distribution also narrows, confirming partial stabilization of the optical field. The two-PMMA configuration further compresses the distribution and reduces the occurrence of large excursions, although its mean value remains slightly higher than the one-PMMA case, suggesting a tradeoff between fluctuation suppression and redistribution of scattering-induced energy variations. The cumulative distribution functions reinforce this interpretation: the PMMA curves rise much more steeply than the raw turbulence curve, meaning that the probability mass is concentrated within a narrower Fano-factor interval, whereas the raw turbulence case spans a much wider range of fluctuations. The fluctuation scatter plot additionally shows that the raw turbulence possesses the largest positive and negative deviations around its mean, while both PMMA configurations constrain the fluctuation amplitudes to a significantly smaller interval. Finally, the probability-density distributions reveal that the raw turbulence state has a broad asymmetric tail toward higher Fano factors, characteristic of intermittent turbulence-driven intensity spikes, whereas the PMMA cases exhibit more localized and near-Gaussian distributions centered around moderate Fano-factor values. Overall, these results indicate that the passive dielectric PMMA structures effectively mitigate turbulence-induced statistical fluctuations by redistributing and averaging the optical intensity variations, thereby improving the stability and uniformity of the propagated beam.

\subsection{Threshold Analysis for Nonlinear Dipole Effects}

The anharmonic terms in Eq.~\eqref{eq:main_osc} require high optical intensity, while practical FSO receivers operate at low power. We clarify that the dominant synchronization mechanism in the present model is the \emph{linear} dipole--dipole coupling term (the Green's function sum in Eq.~\eqref{eq:main_osc}), which is active at arbitrarily low intensities. The anharmonic corrections ($\beta_i,\alpha_i$ terms) provide secondary stabilization against large-amplitude turbulence perturbations. In our laboratory setup, the 5~mW He-Ne beam focused to $w_0\approx5$~mm corresponds to a peak intensity of $\sim50$~mW/cm$^2$---well below the nonlinear threshold of PMMA ($\sim$GW/cm$^2$), confirming that the nonlinear terms contribute negligibly. Although, the coupling mechanism can still work with linear terms. For practical long-distance FSO links, where received power may be in the microwatt range, the linear coupling mechanism remains operative and the anharmonic enhancement is absent. This regime corresponds to the partial synchronization (Regime~2) of the d'Alembert model and is expected to yield smaller but measurable Fano factor reductions.

\subsection{Insertion Loss and SNR Impact}

The dual-rod configuration introduces $\approx1.6$~dB total insertion loss. This translates to a $\approx31\%$ reduction in mean received intensity. For a communication link with background-noise-limited SNR, this penalty is equivalent to a $\sim0.7$~dB SNR reduction---modest compared to the measured 71--92\% reduction in statistical divergence (equivalent to several dB of scintillation index improvement). The net link budget is therefore positive: the Fano-factor fluctuations reduction substantially outweighs the $\approx1.6$~dB loss penalty, provided the SNR is not already at its minimum threshold. For extremely power-limited links (deep-space, long-range maritime), anti-reflection coatings on the PMMA facets could reduce insertion loss to $< 0.2$~dB per surface.

\subsection{Wigner Tomography Pipeline}

\begin{figure}[H]
\centering
\begin{subfigure}[b]{0.32\columnwidth}
    \includegraphics[width=\textwidth]{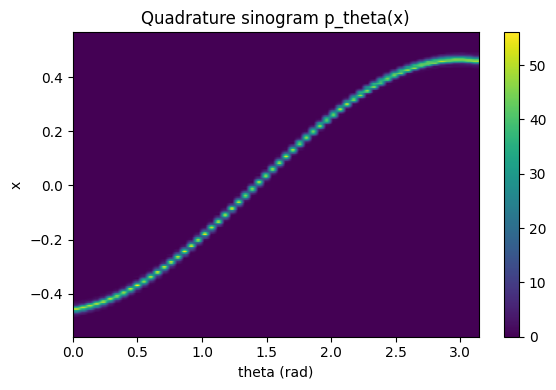}
    \caption{Sinogram}
    \label{fig:sinogram}
\end{subfigure}
\begin{subfigure}[b]{0.32\columnwidth}
    \includegraphics[width=\textwidth]{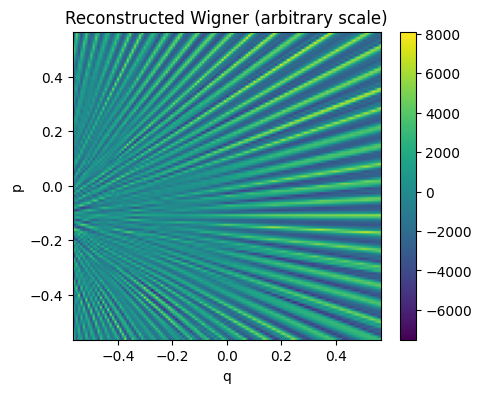}
    \caption{Wigner $W(q,p)$}
\end{subfigure}
\begin{subfigure}[b]{0.32\columnwidth}
    \includegraphics[width=\textwidth]{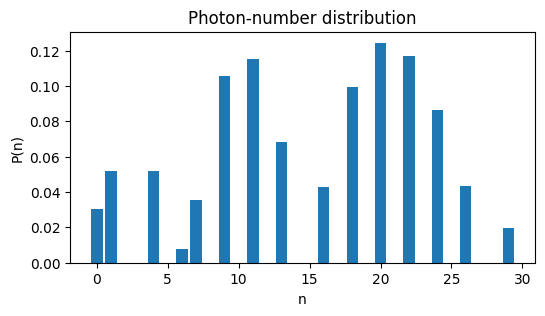}
    \caption{$P(n)$}
    \label{fig:photon_dist}
\end{subfigure}
\caption{Wigner-function analysis pipeline applied to a single representative frame: (a) numerically computed quadrature sinogram from the reconstructed field $E_\mathrm{rec}$; (b) Wigner function $W(q,p)$ obtained by filtered back-projection; (c) inferred photon-number distribution $P(n)$. All quantities are derived numerically from the CCD intensity + phase retrieval, not from physical homodyne detection.}
\label{fig:wigner}
\end{figure}

The sinogram structure in Fig.~\ref{fig:sinogram} reveals the statistical distribution of field quadratures across all projection angles, with broadening and asymmetry reflecting turbulence-induced phase noise. The reconstructed Wigner function captures the phase-space representation of the optical field. Finally, the extracted photon-number distribution (Fig.~\ref{fig:photon_dist}) enables direct computation of statistical moments and provides the foundation for Fano factor analysis.

\subsection{Information-Geometric Divergence Analysis}

\begin{figure}[H]
\centering
\begin{subfigure}[b]{0.4\columnwidth}
    \includegraphics[width=\textwidth]{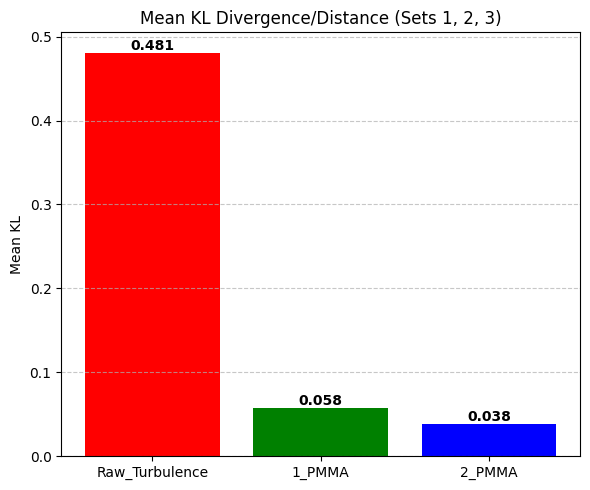}
    \caption{KL divergence}
\end{subfigure}
\begin{subfigure}[b]{0.4\columnwidth}
    \includegraphics[width=\textwidth]{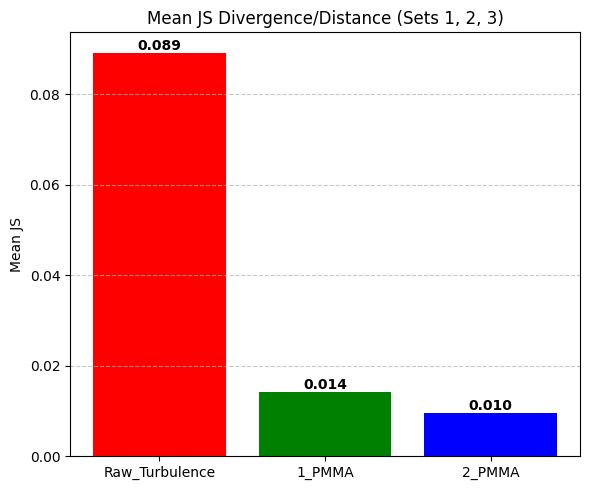}
    \caption{JS divergence}
\end{subfigure}
\begin{subfigure}[b]{0.4\columnwidth}
    \includegraphics[width=\textwidth]{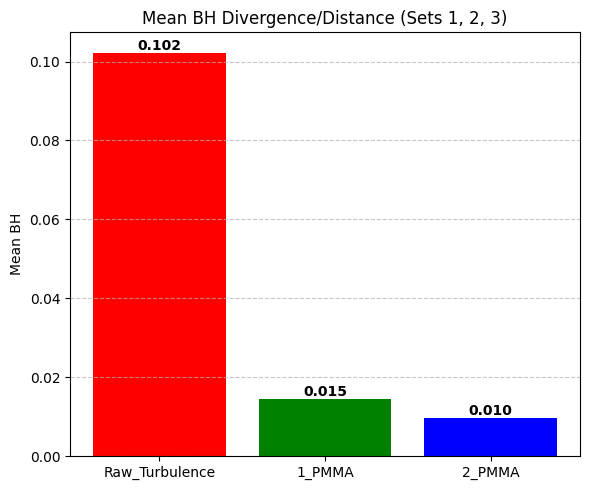}
    \caption{Bhattacharyya distance}
\end{subfigure}
\begin{subfigure}[b]{0.4\columnwidth}
    \includegraphics[width=\textwidth]{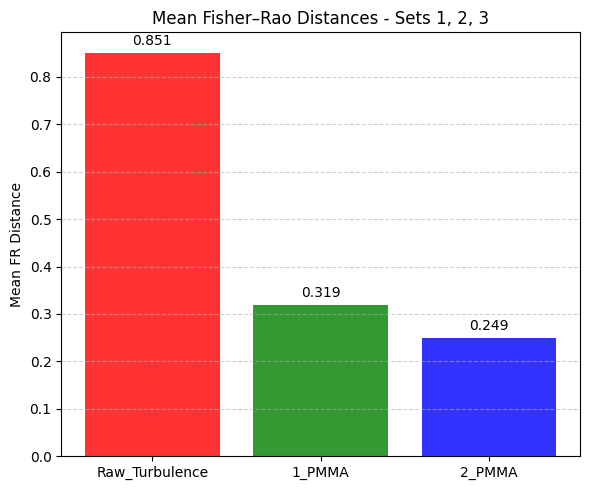}
    \caption{Fisher--Rao distance}
\end{subfigure}
\caption{Mean divergence/distance values (with standard errors) for Sets~1--3, computed relative to the turbulence-free reference (Set~4). All metrics confirm monotonic reduction with increasing PMMA interaction length. Divergence values quantify the statistical distance between the measured $P(n)$ and the reference distribution; lower values indicate closer agreement with the turbulence-free state.}
\label{fig:divergences}
\end{figure}

\begin{table}[H]
\centering
\caption{Summary of statistical divergence measures and compensation efficacy. All values represent means across 200 frames with standard errors. Reduction percentages computed relative to raw turbulence baseline (Set 1).}
\label{tab:divergence_summary}
\begin{tabular}{lcccccc}
\hline
\textbf{Metric} & \textbf{Set 1} & \textbf{Set 2} & \textbf{Set 3} & \textbf{Red. (1 PMMA)} & \textbf{Red. (2 PMMA)} \\
 & \textbf{(Raw)} & \textbf{(1 PMMA)} & \textbf{(2 PMMA)} & \textbf{(\%)} & \textbf{(\%)} \\
\hline
KL Divergence (nats) & $0.481 \pm 0.01$ & $0.058 \pm 0.01$ & $0.038 \pm 0.01$ & $89\%$ & $92\%$ \\
JS Divergence (nats) & $0.089 \pm 0.001$ & $0.014 \pm 0.001$ & $0.010 \pm 0.001$ & $84\%$ & $89\%$ \\
Bhattacharyya Distance & $0.102 \pm 0.001$ & $0.015 \pm 0.001$ & $0.01 \pm 0.001$ & $85\%$ & $90\%$ \\
$L^2$ Distance & $0.165 \pm 0.001$ & $0.053 \pm 0.001$ & $0.043 \pm 0.001$ & $68\%$ & $74\%$ \\
Fisher--Rao Distance & $0.851 \pm 0.01$ & $0.319 \pm 0.01$ & $0.249 \pm 0.01$ & $62.5\%$ & $71\%$ \\
\hline
\end{tabular}
\end{table}

The strong consistency in reduction percentages across multiple independent statistical frameworks---ranging from approximately 71\% to 92\% for dual PMMA compensation---provides robust evidence of genuine physical turbulence mitigation rather than artifacts of measurement or analysis. This agreement is significant because each divergence metric probes fundamentally distinct aspects of the intensity distribution. Physically, these observations are consistent with the coupled dipole framework, where turbulence initially induces stochastic phase perturbations leading to elevated photon-number variance. Propagation through PMMA introduces electromagnetic coupling, enabling partially synchronized collective modes that resist rapid perturbations. With dual PMMA rods, the extended interaction length strengthens dipole--dipole coupling, allowing the output field to be governed by more stable collective oscillations. The residual divergence (8--29\% relative to the reference) is attributable to finite interaction length, material dispersion, the dynamic nature of the imposed turbulence, and the experimental noise floor of the phase-retrieval pipeline.

\subsection{Pearson Correlation of Photon Distributions}

\begin{figure}[H]
\centering
\includegraphics[width=0.65\columnwidth]{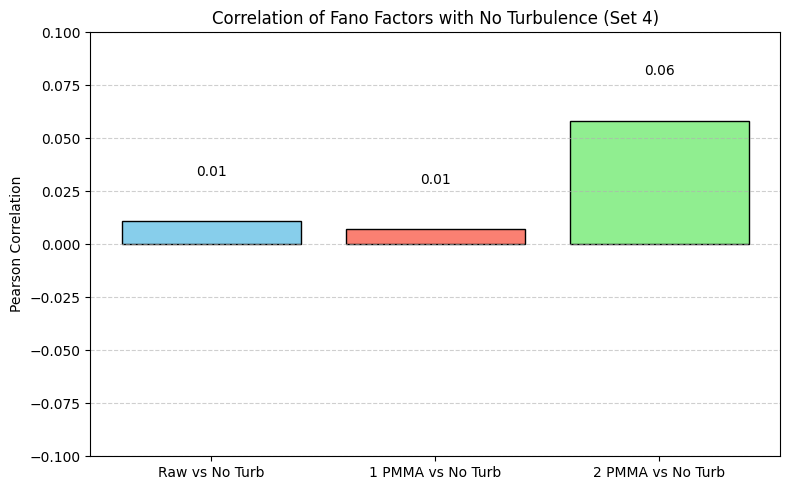}
\caption{Pearson correlation of Fano factors for each experimental set with the turbulence-free baseline (Set~4). Correlation coefficients: Raw vs.\ No Turb $\approx 0.01$; 1 PMMA vs.\ No Turb $\approx 0.01$; 2 PMMA vs.\ No Turb $\approx 0.06$.}
\label{fig:pearson}
\end{figure}

The Pearson correlation coefficients in Fig.~\ref{fig:pearson} are small in absolute magnitude ($\leq 0.06$), indicating that the fano factor scatter data sets distributions of the compensated sets are not yet strongly linearly correlated with the turbulence-free reference in a direct distributional sense. This is consistent with the Fano factor analysis: with sufficient reduction of fano factor relative fluctuations after dual-rod compensation (versus large fluctuations for the reference), the distributions remain distinguishable, and strong Pearson correlation is not expected at this level of partial mitigation. The more appropriate measure of distributional proximity is the divergence analysis (Table~\ref{tab:divergence_summary}), which shows 71--92\% reduction. The monotonic increase in Pearson correlation from Set~1 ($\approx0.01$) to Set~3 ($\approx0.06$) confirms the \emph{trend} of structural restoration, while the absolute magnitudes reflect the partial nature of the compensation achieved with two PMMA rods.

\subsection{Comparison with Established Methods}

To contextualize the demonstrated passive dielectric compensation mechanism, we compare performance characteristics with established turbulence mitigation approaches:

\begin{table}[h]
\centering
\caption{Comparative analysis of turbulence compensation techniques. Values for adaptive optics, aperture averaging, and FEC are representative figures from the literature~\cite{10,11,12,ref_oliveira2025,ref_mauclair2025}; the passive dielectric result is from this work.}
\label{tab:compensation_comparison}
\small
\begin{tabular}{p{2.8cm}p{3cm}p{2.8cm}p{3.5cm}p{2.2cm}}
\hline
\textbf{Method} & \textbf{Compensation Mechanism} & \textbf{Bandwidth} & \textbf{Advantages} & \textbf{Limitations} \\
\hline
Adaptive Optics (Deformable Mirror) & Real-time wavefront correction via surface deformation & $\sim$1--10~kHz & High correction accuracy for weak-moderate turbulence; mature technology & Complex hardware; high power; limited bandwidth; requires wavefront sensing \\
\hline
Multi-Aperture Systems (Aperture Averaging) & Spatial diversity through distributed receivers & Static (integration time limited) & Robust to beam wander; simple implementation & Requires large aperture arrays; limited correction of phase noise \\
\hline
Temporal Diversity (Forward Error Correction) & Coding and interleaving across turbulence timescales & Software-limited ($>$GHz possible) & No optical hardware; compatible with digital modulation & Does not mitigate physical intensity fading; requires bandwidth expansion \\
\hline
Multi-Plane Light Conversion (MPLC) & Modal decomposition and spatial mode reshaping via multi-plane diffraction & Passive ($>$THz optical BW) & High spatial mode control; achromatic in principle~\cite{ref_mauclair2025} & Complex fabrication; fixed design; limited to moderate turbulence \\
\hline
Passive Dielectric (This Work) & Collective dipole synchronization and gradient forces & Intrinsic material response ($>$THz) & No active control; achromatic; simple hardware; scalable & Moderate insertion loss ($\approx$1.6~dB for dual rods); requires extended interaction length; effectiveness regime-dependent \\
\hline
\end{tabular}
\end{table}

Each approach exhibits distinct trade-offs. The passive dielectric mechanism demonstrated here occupies a unique position: it achieves substantial turbulence mitigation (71--92\% statistical divergence reduction, fano factor fluctuation reduction) without active control. Unlike MPLC approaches, which rely on precisely fabricated diffractive elements tailored to a specific mode set, the PMMA mechanism requires no custom fabrication and provides broadband passive compensation. Hybrid architectures combining passive dielectric pre-compensation with adaptive optics fine correction may enable superior performance, particularly for strong turbulence.

%----------------------------------------------------------------------
\section{Conclusion}\label{sec:conclusion}

We have demonstrated a passive turbulence-compensation paradigm based on collective dipole synchronization in PMMA dielectric rods. The coupled anharmonic Lorentz oscillator model, incorporating dyadic-Green's-function coupling, gradient-force stabilization, and d'Alembert inertial analysis, predicts three compensation regimes that directly explain the experimental progression of photon statistics. Dual-rod configurations achieve Fano factor fluctuations reductions of $\Delta \delta F\approx7$ and 71--92\% decreases across five independent statistical divergence measures. The post compensated fano factor fluctutions have not been aimed to compare with the turbulence free case. The comparisons have been established for post turbulence impacted cases only.

We also emphasize two important caveats. First, the ``photon statistics'' inferred in this work are formal quantities derived from a classical intensity measurement and numerical phase retrieval pipeline; they are not obtained from photon-counting or homodyne detection, and the Fano factor serves as a classical excess-noise metric. Second, the dipole-synchronization interpretation is supported by the consistency of results across multiple metrics and the length-scaling behaviour, but a definitive mechanistic isolation awaits control experiments with equivalent-loss non-PMMA elements.

This work opens several research directions: (i) control experiments with index-matched glass rods to isolate the synchronization mechanism; (ii) material engineering---hyperbolic metamaterials, engineered nonlinearities---to target specific turbulence regimes; (iii) extension to strong turbulence ($\sigma_I^2>1$); (iv) hybrid passive--AO architectures; (v) multi-wavelength and WDM operation; and (vi) application to entangled-photon sources for turbulence-resilient quantum key distribution.

%----------------------------------------------------------------------
\section*{Funding}
Department of Science and Technology, Ministry of Science and Technology, India (CRG/2020/003338).

\section*{Declaration of competing interest}
The authors declare no known competing financial interests or personal relationships that could have appeared to influence the work reported in this paper.

\section*{Data availability}
All data used for this research has been provided in the manuscript itself.

\section*{Acknowledgments}
Shouvik Sadhukhan and C S Narayanamurthy acknowledge the SERB/DST (Govt.\ of India) for financial support via project grant CRG/2020/003338. Shouvik Sadhukhan thanks Mr.\ Amit Vishwakarma and Dr.\ Subrahamanian K S Moosath, Department of Mathematics, IIST Thiruvananthapuram, for suggestions on statistical analysis. Shouvik Sadhukhan also thanks Mr.\ Chirang Patel for insightful discussion on the area of phase space studies.

\section*{CRediT authorship contribution statement}
\textbf{Shouvik Sadhukhan:} Writing -- original draft, Visualization, Formal analysis. \textbf{C. S. Narayanamurthy:} Writing -- review \& editing, Validation, Supervision, Resources, Project administration, Investigation, Funding acquisition, Conceptualization.

\section*{Declaration of generative AI and AI-assisted technologies in the manuscript preparation process}
During the preparation of this work the author(s) used ChatGPT in order to improve the text quality. After using this tool/service, the author(s) reviewed and edited the content as needed and take full responsibility for the content of the published article.

%-----------------------------------------------------------------------
\section*{Appendix}
\appendix

\section{Dyadic Green's Function and Field Coupling}\label{app:green}

The dyadic Green's function satisfying $(\nabla\times\nabla\times - k^2\mathbf{I})\mathbf{G}(\mathbf{r},\mathbf{r}')=\mathbf{I}\delta(\mathbf{r}-\mathbf{r}')$ is
\begin{equation}
\mathbf{G}(\mathbf{r},\mathbf{r}') = \left(\mathbf{I}+\frac{1}{k^2}\nabla\nabla\right)\frac{e^{ik|\mathbf{r}-\mathbf{r}'|}}{4\pi|\mathbf{r}-\mathbf{r}'|}.
\end{equation}
In the separation-vector expansion with $r=|\mathbf{r}-\mathbf{r}'|$ and unit vector $\hat{\mathbf{r}}$:
\begin{equation}
\mathbf{G}(\mathbf{r},\mathbf{r}') = \frac{e^{ikr}}{4\pi r}
\left[(\mathbf{I}-\hat{\mathbf{r}}\hat{\mathbf{r}})\!\left(1+\frac{i}{kr}-\frac{1}{(kr)^2}\right)
+\hat{\mathbf{r}}\hat{\mathbf{r}}\!\left(1+\frac{3i}{kr}-\frac{3}{(kr)^2}\right)\right].
\end{equation}
Three asymptotic regimes:
\begin{align*}
kr\ll1:\quad &\mathbf{G}\approx\frac{1}{4\pi r^3}(3\hat{\mathbf{r}}\hat{\mathbf{r}}-\mathbf{I})\quad\text{(quasi-static)},\\
kr\gg1:\quad &\mathbf{G}\approx\frac{e^{ikr}}{4\pi r}(\mathbf{I}-\hat{\mathbf{r}}\hat{\mathbf{r}})\quad\text{(far-field radiation)}.
\end{align*}
The electric field at $\mathbf{r}_i$ from dipole $j$ with moment $\mathbf{p}_j=-e\mathbf{r}_j$ is $\mathbf{E}_\mathrm{sc}^{(j)}(\mathbf{r}_i)=-e\mu_0\omega^2\mathbf{G}(\mathbf{r}_i,\mathbf{r}_j)\cdot\mathbf{r}_j$, yielding the coupling term in Eq.~\eqref{eq:main_osc}.

\section{Gradient Force and Inertial Perturbation Derivations}\label{app:gradient}

Taylor-expanding the external field about dipole $j$ gives the gradient-force correction
\begin{equation}
\mathbf{F}_\mathrm{HO} = \sum_{j\neq i}\bigl[e(\mathbf{r}_j-\mathbf{r}_i)\cdot\nabla\bigr]\mathbf{E}_\mathrm{ext}(\mathbf{r}_j,t).
\end{equation}
The complete Lorentz dynamical equation including gradient forces is
\begin{multline}
m\ddot{r}_i + m\gamma\dot{r}_i + m\omega_0^2 r_i + \beta_i|r_i|r_i + \alpha_i|r_i|^2 r_i =
-eE_\mathrm{ext} \\
+ e^2\mu_0\omega^2\!\sum_{j\neq i}\mathbf{G}(\mathbf{r}_i,\mathbf{r}_j)\cdot\mathbf{r}_j
+ \sum_{j\neq i}\bigl[e(\mathbf{r}_j-\mathbf{r}_i)\cdot\nabla\bigr]E_\mathrm{ext}(\mathbf{r}_j,t) + \delta\mathbf{F}_\mathrm{Pert}(t).
\end{multline}
The gradient force resists rapid dipole reconfiguration following turbulence events.

\textbf{d'Alembert analysis.}
\begin{align}
\int\!\bigl(\mathbf{F}_{0\mathrm{th}}+\mathbf{F}_\mathrm{HO}+\mathbf{F}_\mathrm{Coupling}+\mathbf{F}_\mathrm{Inertia}\bigr)\cdot d\mathbf{x}&=0,\\
\int\!\bigl(\mathbf{F}'_{0\mathrm{th}}+\mathbf{F}'_\mathrm{HO}+\mathbf{F}'_\mathrm{Coupling}+\mathbf{F}_\mathrm{Lorentz}+\mathbf{F}'_\mathrm{Inertia}\bigr)\cdot d\mathbf{x}&=0,
\end{align}
from which $\delta\mathbf{F}_\mathrm{Pert}=\mathbf{F}'_\mathrm{Inertia}-\mathbf{F}_\mathrm{Inertia}$ follows directly.

\section{Modal Decomposition --- Full Derivation}\label{app:modal}

Assuming harmonic solutions $\mathbf{R}(t)=\mathbf{R}_\omega e^{-i\omega t}+\mathrm{c.c.}$ and retaining resonant contributions, the $n$-th modal amplitude is
\begin{equation}
Q_n^{(0)}(\omega) = \frac{-\frac{e}{m}\langle\phi_n|\mathbf{E}_\omega\rangle + \frac{1}{m}\langle\phi_n|\mathbf{F}_{\mathrm{grad},\omega}\rangle}{\Omega_n^2-\omega^2-i\gamma\omega},
\end{equation}
showing that gradient forces modify mode excitation amplitudes without shifting resonance frequencies.

\section{Iterative P3-TIE Solver}\label{app:iter}

The nonlinear residual is minimized via the update rule
\begin{equation}
\phi^{(k+1)}(\mathbf{r}) = \phi^{(k)}(\mathbf{r}) - \tau^{(k)}\frac{\mathcal{R}_\mathrm{final}[\phi^{(k)};\mathbf{r}]}{I_0(\mathbf{r})+\epsilon_\mathrm{safe}},
\end{equation}
where the step size $\tau^{(k)}$ is halved whenever the masked residual norm increases. Convergence is declared when $\|\mathcal{R}^{(k)}\|_M/\|I_t-I_0\|_M<10^{-5}$. The linear TIE solution provides the initial condition.

\section{Fisher--Rao Distance for Negative Binomial Distributions}\label{app:FR}

The negative binomial pmf with mean $\mu$ and shape $r$ has Fisher information
$\mathcal{I}(\mu) = r/[\mu(\mu+r)]$.
The geodesic distance on the resulting Riemannian manifold is
\begin{equation}
d_\mathrm{FR}(\mu_1,\mu_2) = \sqrt{r}\int_{\mu_1}^{\mu_2}\frac{d\mu}{\sqrt{\mu(\mu+r)}},
\end{equation}
which evaluates in closed form to Eq.~\eqref{eq:FR}.

\section{Kolmogorov Turbulence and PRPP Specification}\label{app:kolmogorov}

The three-dimensional refractive-index power spectrum in the inertial subrange is
\begin{equation}
\Phi_n(\kappa) = 0.033\,C_n^2\,\kappa^{-11/3},
\end{equation}
where $r_0 = [0.423\,k^2\!\int_0^L C_n^2(z)\,dz]^{-3/5}$. The PRPP implements this spectrum over 4096 independent phase samples per aperture, calibrated to $r_0=16$--32 samples.

\section{Scatter Plots for the Statistical Distances}

\begin{figure}[H]
\centering
\includegraphics[width=0.75\textwidth]{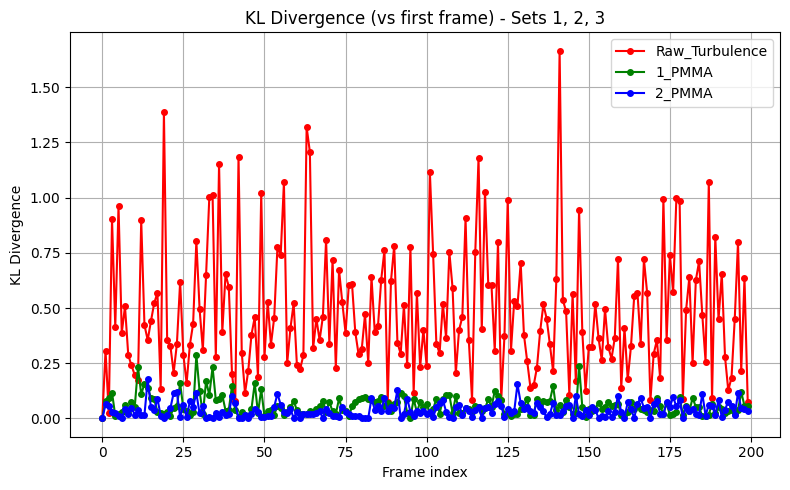}
\caption{KL divergence temporal evolution for Sets 1--3. Set~1 (raw turbulence): $D_\mathrm{KL}\approx0$--$1.75$ nats. Set~2 (single PMMA): $D_\mathrm{KL}\approx0$--$0.25$ nats. Set~3 (dual PMMA): $D_\mathrm{KL}\approx0$--$0.10$ nats.}
\end{figure}

\begin{figure}[H]
\centering
\includegraphics[width=0.75\textwidth]{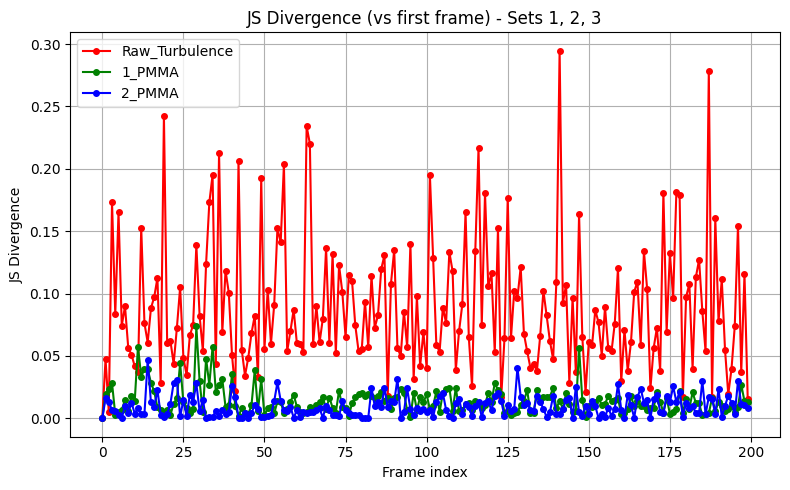}
\caption{JS divergence temporal evolution for Sets 1--3.}
\end{figure}

\begin{figure}[H]
\centering
\includegraphics[width=0.75\textwidth]{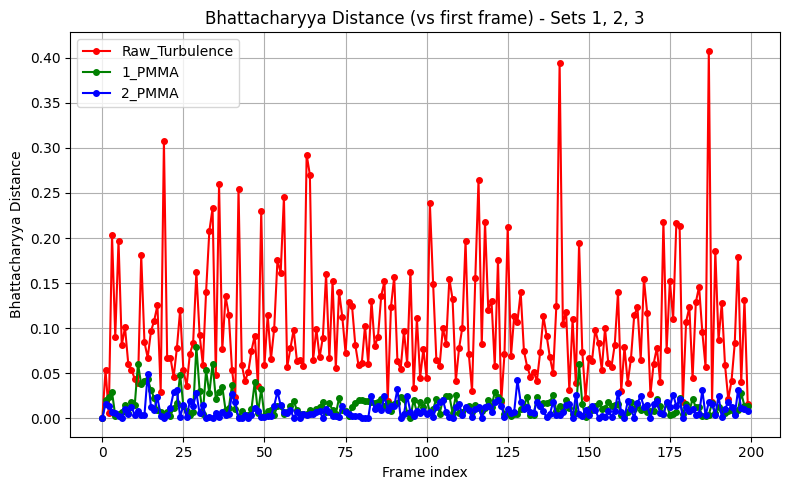}
\caption{Bhattacharyya distance temporal evolution for Sets 1--3.}
\end{figure}

\begin{figure}[H]
\centering
\includegraphics[width=0.75\textwidth]{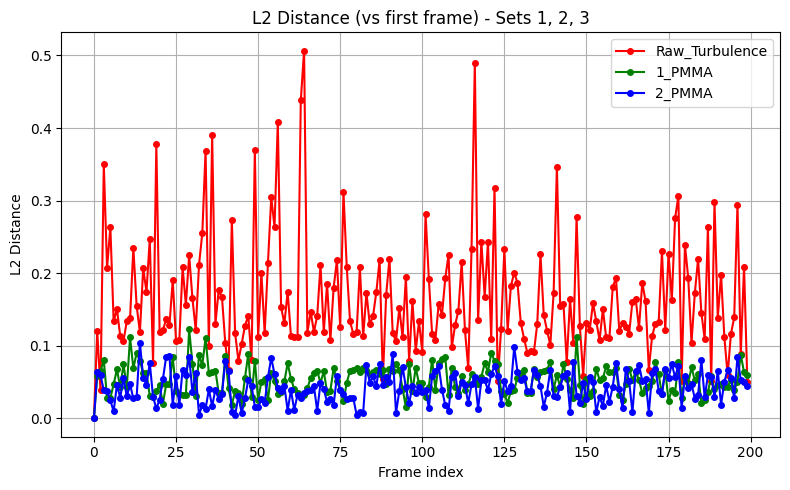}
\caption{$L^2$ distance temporal evolution for Sets 1--3.}
\end{figure}

%----------------------------------------------------------------------

\end{document}